\documentclass[useAMS,usenatbib]{mn2e}

\usepackage{epsfig,color}
\usepackage{graphicx}
\usepackage[]{txfonts}
\usepackage{amssymb}
\usepackage{subfig}
\usepackage{tabularx}

\newcommand{\eqb}{\begin{eqnarray}}
\newcommand{\eqe}{\end{eqnarray}}
\newcommand{\sth}{\sigma_{\rm T}}

\usepackage{aas_macros}

\title{A combined radio and GeV $\gamma$-ray view of the 2012 and 2013 flares of Mrk~421}

\author[T. Hovatta et al.]
{\parbox{\textwidth}{T. Hovatta$^{1,2}$\thanks{E-mail: talvikki.hovatta@aalto.fi},
M. Petropoulou$^{3}$\thanks{Einstein Postdoctoral Fellow},
J. L. Richards$^{3}$,
D. Giannios$^{3}$,
K. Wiik$^{4}$,
M. Balokovi{\' c}$^{2}$,
A. L\"ahteenm\"aki$^{1,5}$,
B. Lott$^{6,7}$,
W. Max-Moerbeck$^{8}$,
V. Ramakrishnan$^{1}$,
A. C. S. Readhead$^{2}$}\vspace{0.4cm}\\
\parbox{\textwidth}{
$^{1}$Aalto University Mets\"ahovi Radio Observatory, Mets\"ahovintie
114, FI-02540 Kylm\"al\"a, Finland\\
$^{2}$Cahill Center for Astronomy and
  Astrophysics, California Institute of Technology, Pasadena CA,
  91125, USA\\
$^{3}$Department of Physics, Purdue University, 525 Northwestern Ave, West Lafayette, IN 47907, USA\\
$^{4}$Tuorla Observatory, Department of Physics and Astronomy,
University of Turku, V\"ais\"al\"antie 20, FI-21500 Piikki\"o, Finland\\
$^{5}$Aalto University Department of Radio Science and Engineering, 13000, FI-00076 AALTO, Finland\\
$^{6}$Univ. Bordeaux, CENBG, UMR 5797, F-33170 Gradignan, France\\
$^{7}$CNRS, IN2P3, CENBG, UMR 5797, F-33170 Gradignan, France\\
$^{8}$National Radio Astronomy Observatory, P.O. Box 0, Socorro, NM 87801, USA.\\
}}
\begin{document}

\date{Accepted XXX. Received YYY; in original form ZZZ}

\pagerange{\pageref{firstpage}--\pageref{lastpage}} \pubyear{2014}

\maketitle

\label{firstpage}

\begin{abstract}
  In 2012 Markarian~421 underwent the largest flare ever observed in
  this blazar at radio frequencies.   In the present study, we
    start exploring this unique event and compare it to a less extreme event
    in 2013. We use 15\,GHz radio data obtained with the Owens Valley
  Radio Observatory 40-m telescope, 95\,GHz millimeter data from the
  Combined Array for Research in Millimeter-Wave Astronomy, and GeV
  $\gamma$-ray data from the {\it Fermi Gamma-ray Space Telescope}.
   The radio light curves during the flaring periods in 2012 and 2013
  have very different appearances, both in shape and peak flux
  density.  Assuming that the radio and $\gamma$-ray flares are
  physically connected, we attempt to model the most prominent
  sub-flares of the 2012 and 2013 activity periods by using the
  simplest possible theoretical framework. We first fit a one-zone
  synchrotron self-Compton (SSC) model to the less extreme 2013 flare
  and estimate parameters describing the emission region. We then
  model the major $\gamma$-ray and radio flares of 2012 using the same
  framework. The 2012 $\gamma$-ray flare shows two distinct spikes of
  similar amplitude, so we examine scenarios associating the radio
  flare with each spike in turn. In the first scenario, we cannot
  explain the sharp radio flare with a simple SSC model, but we can
  accommodate this by adding plausible time variations to the Doppler
  beaming factor.  In the second scenario, a varying Doppler factor is
  not needed, but the SSC model parameters require fine tuning.  Both
  alternatives indicate that the sharp radio flare, if physically
  connected to the preceding $\gamma$-ray flares, can be reproduced
  only for a very specific choice of parameters.
\end{abstract}

\begin{keywords}
galaxies: active -- galaxies: jets -- BL Lacertae objects: individual:
Markarian 421

\end{keywords}

\section{Introduction}
Blazars are active galactic nuclei (AGN) that are variable over the
entire electromagnetic spectrum. Their variability is enhanced due to
Doppler-beamed emission from a relativistic jet pointing close to the
line of sight. Blazars having featureless optical spectra or showing emission lines with
equivalent width $< 5$~{\AA} are historically classified as BL Lacs
\citep{stocke91}.

Markarian 421 (hereafter Mrk~421) is one of the best studied BL~Lac
 objects. It is relatively nearby with a redshift of $z=0.031$ \citep{ulrich75}.
It is classified as a high synchrotron peaked blazar \citep{abdo10c} based on its
 spectral energy distribution (SED). It was the
 first blazar detected at TeV energies
 \citep{punch92}, and since then it has been the subject of numerous multi-wavelength
 campaigns \citep[e.g.,][]{aleksic12,abdo11,acciari11,horan09,donnarumma09,fossati08,rebillot06,
 tosti98}. The broad-band SED of Mrk~421 can often be modeled
with a one-zone synchrotron self-Compton (SSC) model \citep{aleksic12,
  acciari11, abdo11} or with lepto-hadronic models involving proton
synchrotron radiation and/or
photopion interactions \citep[e.g.,][]{bottcher13, mastichiadis13, dimitrakoudis14}.

Mrk~421 exhibits large variations in the TeV,
GeV, X-ray, and optical wavebands
\citep[e.g.,][]{acciari11,acciari09,horan09,blazejowski05}, with
correlated TeV and X-ray variations
\citep[e.g.,][]{giebels07,fossati08}. The connection between the TeV and
optical bands is less clear and more detailed time-dependent models
are being developed to study the complex correlations (see e.g.,
\citealt{chen11} for SSC modeling and \citealt{mastichiadis13} for
lepto-hadronic modeling).

In 2012 Mrk~421 underwent its largest radio flare ever observed at
15\,GHz \citep{hovatta12ATel}. The flare time-scale  was
  faster and its amplitude larger than 
  any other radio flare observed 
  from this source in the past 30 years of observations with the University of
  Michigan Radio Astronomy Observatory monitoring program
  \citep{richards13}.
The radio flare occurred about 40~d
after a major flare was observed in the $\gamma$-ray band
\citep{Dammando12ATel} by the {\it Fermi Gamma-Ray Space Telescope} (hereafter {\it
Fermi}). Unfortunately Mrk~421 was close to the sun at the
time of the radio flare so that only limited multi-wavelength coverage in
addition to the radio and $\gamma$-ray bands exists, apart from a
NuSTAR calibration observation that covered the early activity \citep{balokovic13b}. In the spring of 2013,
Mrk~421 underwent another major high-energy flaring event  in the
  X-ray to TeV energies \citep{cortina13,balokovic13,paneque13}. About 60~d later, it was
followed by fairly small amplitude radio flares in the radio and
millimeter bands \citep{hovatta13ATel}.

In this paper we present the radio data obtained with the Owens Valley
Radio Observatory (OVRO) 40-m telescope at 15\,GHz, and 95\,GHz data
obtained with Combined Array for Research in Millimeter-Wave Astronomy
(CARMA). We compare the radio variations with the $\gamma$-ray light
curves obtained by {\it Fermi}, by performing a cross-correlation
analysis on the full data sets available from 2008 to 2013. Based on
our findings and the coincidence of historically rare
  extreme flares in the two bands, we assume that the events in the
radio and $\gamma$-ray bands are physically connected. Under this
  assumption we aim to understand the extreme nature of the 2012 radio
  event by adopting a reasonable physical model with the smallest
  number of free parameters.

Our paper is organized as follows. We describe
the observations and data reduction in Sect.~\ref{sect:obs}. The light curves and
cross-correlations between the different bands are shown in
Sect.~\ref{sect:lc}. We model the flares in Sect.~\ref{sect:model}, discuss our results in
Sect.~\ref{sect:discussion}, and present our conclusions in Sect.~\ref{sect:conclusions}.
Throughout the paper we use a cosmology where $H_0 = 71~\mathrm{km}~\mathrm{s}^{-1}~\mathrm{Mpc}^{-1}$,
$\Omega_M = 0.3$, and $\Omega_\Lambda = 0.7$
\citep[e.g.,][]{komatsu09}.

\section{Observations}\label{sect:obs}
We present radio and $\gamma$-ray observations of the two flaring
  events observed in 2012 and 2013. We are especially
    concerned with the extreme 2012 radio event, whereas the 2013 flare is
    more typical of the radio variability observed in Mrk~421. 

\subsection{Radio observations}
Mrk~421 was observed as part of the blazar monitoring program\footnote{http://www.astro.caltech.edu/ovroblazars/} with the
OVRO 40-m telescope \citep{richards11}. In this program, a sample of
over 1800 AGN are observed twice per week at 15\,GHz. Mrk~421 has been
included since the beginning of the monitoring in late 2007.

The OVRO 40-m telescope uses off-axis dual-beam optics and a cryogenic high
electron mobility transistor (HEMT) low-noise amplifier with a
15.0~GHz center frequency and 3~GHz bandwidth. The two sky beams are Dicke switched using the
off-source beam as a reference, and the source is alternated between
the two beams in an ON-ON fashion to remove atmospheric and ground
contamination. A noise level of approximately 3--4~mJy in quadrature
with about 2 per cent additional uncertainty, mostly due to pointing errors,
is achieved in a 70~s integration period. Calibration is achieved
using a temperature-stable diode noise source to remove receiver gain
drifts and the flux density scale is derived from observations of
3C~286 assuming the \cite{baars77} value of 3.44~Jy at
15.0~GHz. The systematic uncertainty of about 5 per cent in the flux density
scale is not included in the error bars.  Complete details of the
reduction and calibration procedure are given in Richards et
al. (2011). 

For comparison and to fill gaps in the OVRO sampling,
  we also  consider the 37 GHz light curve obtained by the Mets\"ahovi
  Radio Observatory.  These observations were made with the
  13.7-m radome-enclosed telescope using a 1\,GHz-bandwidth
  dual beam receiver centered at 36.8\,GHz. A detailed description of
  the data reduction and analysis is given in \cite{terasranta98}.  As
  the uncertainty in the data points is much larger than in the other
  bands, we do not include the data in any subsequent modeling. 

\subsection{Millimeter-band observations}
Since February 2013, Mrk~421 was observed $1-3$ times per week with
CARMA as a part of the MARMOT\footnote{Monitoring of
    $\gamma$-ray Active galactic nuclei with Radio, Millimeter and
    Optical Telescopes; http://www.astro.caltech.edu/marmot/} blazar
  monitoring project. The observations were made using the
eight 3.5~m telescopes of the array with a central frequency of
95\,GHz and a bandwidth of 7.5\,GHz. The data were reduced
using the MIRIAD (Multichannel Image Reconstruction, Image Analysis
and Display) software \citep{sault95}, including standard
bandpass calibration on a bright quasar.  The amplitude and phase
  gain calibration was done 
by self-calibrating on Mrk~421. The absolute flux calibration was
determined from a  temporally nearby observation (within a day) of the planets
Mars, Neptune or Uranus, whenever possible. Otherwise the quasar
3C~273 was used as a secondary calibrator. The estimated absolute
calibration uncertainty of 10 per cent is not included in the error
bars.

\subsection{Gamma-ray data}
The $\gamma$-ray data were obtained with the Large Area Telescope
(LAT) aboard {\it Fermi}, which observes the entire sky every 3 hours at energies of
$0.1-300$\,GeV \citep{atwood09}. The publicly available reprocessed Pass
7 data\footnote{http://fermi.gsfc.nasa.gov/ssc/data/} were downloaded and analysed using the {\it Fermi} ScienceTools software
package version v9r32p5. The data were binned using the adaptive binning method of
\cite{lott12}. The uneven bin size was determined in such a way that
the statistical error
in each flux measurement is $\sim$15 per cent
We used the instrument response functions
P7REP\_SOURCE\_V15, Galactic diffuse emission model ``gll\_iem\_v05.fits''  and
isotropic background model ``iso\_source\_v05.txt''.\footnote{http://fermi.gsfc.nasa.gov/ssc/data/access/lat/BackgroundModels.html} Source class
  photons (evclass=2) within 15$^\circ$ of Mrk~421 were selected, with
  a zenith angle cut of 100$^\circ$ and a rocking angle cut of 52$^\circ$.

Once the bins were defined, the photon fluxes in the energy range of 0.1 -
200\,GeV were calculated using unbinned likelihood
and the tool {\it gtlike} with the Minuit optimizer. All sources within
15$^\circ$ of Mrk~421 were included in the
source model with their spectral parameters, except the flux, frozen to the values determined
in the 2nd {\it Fermi} LAT catalog (2FGL; \citealt{nolan12}). For sources more
than 10$^\circ$ from Mrk~421 we also froze the fluxes to the 2FGL
value. The 10 per cent systematic uncertainty below 100 MeV, decreasing
  linearly in Log(E) to 5 per cent in the range between 316~MeV and 10~GeV
  and increasing linearly in Log(E) up to 15 per cent at 1~TeV  \citep{ackermann12c} is not included in the error bars.
For the purpose of the modeling presented in
  Sect.~\ref{sect:model}, we also obtained the energy fluxes, with a power-law spectral
  model where the index is frozen to the 2FGL catalog value of
  $\Gamma=1.77$.

\section{Light curves}\label{sect:lc}
\begin{figure}
\includegraphics[scale=0.4]{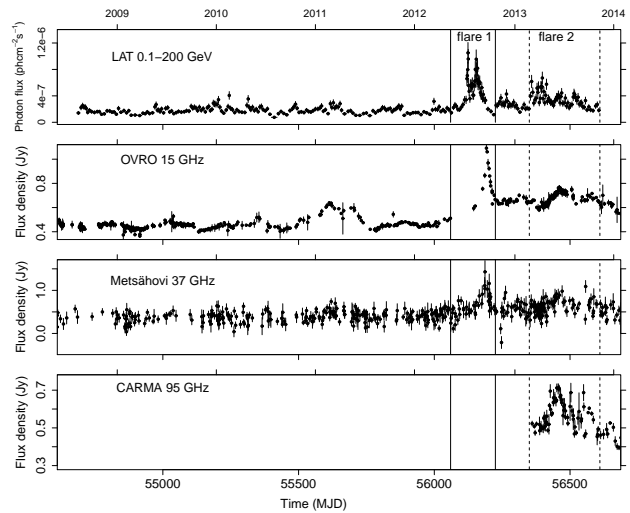}
\caption{From top to bottom: light curves of Mrk~421 in $\gamma$ rays from {\it Fermi} LAT, radio 15\,GHz from OVRO, 37\,GHz from Mets\"ahovi, and 95\,GHz from CARMA. The light curves span a time period from August 2008 until
November 2013, except for the CARMA light curve that starts in February 2013. The solid lines indicate the time range used to model
the 2012 flare from 13 May 2012 (MJD 56060) until 25 October 2012 (MJD
56225). The dashed lines indicate the time range for the 2013 flare
from 9 March 2013 (MJD 56360) until 14 November 2013 (MJD
56610). Because of the larger uncertainties in the data, the
  37\,GHz light curve is not used in any subsequent modeling.}\label{fig:lc}
\end{figure}

The radio, millimeter and $\gamma$-ray light curves are shown in
Fig.~\ref{fig:lc}. They cover the time range since the beginning of
the {\it Fermi} mission in August 2008 (MJD~54688) until the end of
October 2013 (MJD~56610).  The light curves illustrate the unusual nature of the 2012
flare (flare 1) in both radio and $\gamma$ rays, that lasted from May
2012 until October 2012 (MJD 56060 - 56225). This is a unique
  event, especially in the radio band where such fast and prominent
  flares have not been observed before in Mrk~421.  The 2013 flare
(flare 2) from March 2013 until November 2013 (MJD 56350-56610), although prominent, is
much broader and of lower amplitude.

A full cross-correlation analysis between more than four years of OVRO and
LAT data was done by \cite{max-moerbeck14}. They used weekly binned
$\gamma$-ray light curves and data from 2008 until November 2012,
including the rapid 2012 flare. They found a peak in the
discrete correlation function (DCF), with the $\gamma$ rays leading
the radio by $40\pm9$~d. The significance of this correlation was
between 96.16 and 99.99 per cent depending on the power spectral density (PSD) model used for the
light curves, with a best-fitting value of 98.96 per cent. For details of the significance estimation, see
  \cite{max-moerbeck14a}.
We repeated the cross-correlation
  analysis using the extended light curves  considered here, and found that the DCF
  shows
 a broad peak   ($\sim$ 30~d). In particular, the time delay ranges from 40 to  70~d with $\gamma$ rays leading,
  consistent with the estimate from \cite{max-moerbeck14}. The
  significance of the peak was from 91.90 to 99.99 per cent, with a best fit
  value of 97.36 per cent, depending on the PSD model.  The difference
    compared to the exact value derived by \cite{max-moerbeck14} is
    because they only considered the  maximum of the
     DCF, while the peak itself is broad.

  Recently, \cite{emmanoulopoulos13}
  introduced a method for simulating light curves, which also accounts
  for  their flux distribution. This is more appropriate in the $\gamma$-ray
  light curves, as the light curves  have a non-Gaussian photon
    flux distribution. Using this
  method, the significance of the correlation increases to 99.82 per cent, when
  using the best-fitting PSD. In the rest of the paper, we will  thus assume that the major flares in radio and
$\gamma$ rays are physically connected. 
  
 The $\gamma$-ray flares in both 2012 and 2013 have
  significant sub-structure, as already shown in Fig.~\ref{fig:lc}
  (top panel).  If we also consider the broadness of the DCF peak, it
  becomes unclear which $\gamma$-ray spike of the overall flare is
   most plausibly associated with the radio
  peaks. Therefore, we  consider both alternatives in our
  modeling.

\subsection{The flare of 2012}\label{sect:2012}

On 16 July 2012 (MJD 56124) the $\gamma$-ray flux reached the highest
value since the start of the {\it Fermi} mission \citep{Dammando12ATel}. The
$\gamma$-ray flux increased by a factor of three from  $(3.4 \pm 0.6)
\times10^{-7}$\,ph\,cm$^{-2}$\,s$^{-1}$ to $(1.1 \pm 0.2) \times10^{-6}$\,ph\,cm$^{-2}$\,s$^{-1}$ within seven days. The flare
appears double peaked with the second flare peaking on 15 August 2012 (MJD 56154) at a
flux of $(9.5 \pm 0.1)\times10^{-7}$\,ph\,cm$^{-2}$\,s$^{-1}$. The total duration of the flaring event was about
 60~d.  The OVRO 15\,GHz light curve exhibits a fast rise which leads
to an increase of the flux density by a factor of about two, i.e. from 0.6 to 1.1 Jy, in the period
 6 August 2012 -- 21 September 2012.
This flux density is higher than any flux density measured at 15\,GHz
 in the OVRO program or  during the preceding 30-plus years of monitoring with the University of Michigan Radio
Astronomy Observatory \citep{richards13}.

 At the time of the first $\gamma$-ray peak there is a gap in the
  OVRO 15\,GHz light curve.  Thus, a double peaked radio flare cannot
  be excluded {\sl a priori}. However, a higher frequency radio light
  curve observed at 37\,GHz at Mets\"ahovi Radio Observatory (third
  panel from the top in Fig.~\ref{fig:lc}) shows  only a
  single flare during 2012. Given the  spectral proximity of the two radio
  bands  and the similar features in the two light curves, it is safe to assume
that the OVRO sampling does not miss a peak at the beginning of the
event. 
Because of the larger statistical uncertainties in the 37\,GHz
data, we do not include them in  our subsequent analysis as they would
not add additional constraints on the modeling. 

\begin{figure}
\includegraphics[scale=0.52]{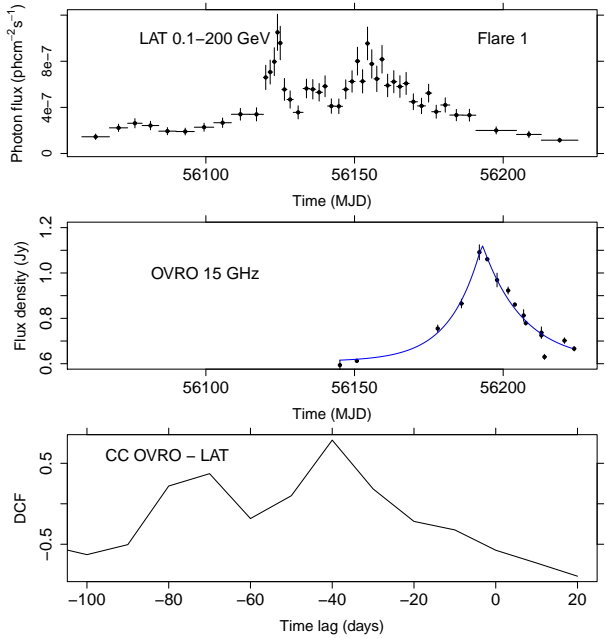}
\caption{$\gamma$-ray flux (top) and 15\,GHz flux density (middle)
  during the 2012 flare.  The discrete correlation function between
  the two light curves is shown in the bottom panel. A negative time
  delay means that $\gamma$ rays are leading the radio. The (blue)
  solid line shows the exponential fit that is used to estimate the
  variability  brightness temperature and Doppler factor. The residuals
  of the fit are on average less than 20~mJy. The time range in the
  figure corresponds to the solid lines in
  Fig.~\ref{fig:lc}}.\label{fig:cc1}
\end{figure}

In order to obtain a better estimate of the time delay between the
$\gamma$-ray and 15\,GHz light curves for this flare only, we use the
DCF method over the time period of the flare ( Fig.~\ref{fig:cc1}, bottom panel).  There are two peaks in the DCF which
simply correspond to the delays between each of the  spikes in
the double-peaked $\gamma$-ray flare and the single radio 
  flare.  The DCF peaks are both consistent with the
  delay measurement we obtained for the full light curves.  
  Although the correlation is stronger for the  $\sim$40~d lag, the
  amplitude of the DCF peak at  about  $-70$~d is not low
   enough to justify exclusion of a possible association of
  the radio flare with first $\gamma$-ray spike.  Thus, we will test
  both possibilities in Sect.~\ref{sect:model}.

We can estimate the Doppler boosting factor of the radio flare by assuming
that the rise time of the flare corresponds to the light travel time
across the emission region \citep{lahteenmaki99b, hovatta09}. We fit
 an exponential function of the form
\begin{equation}
S(t) = \Delta S e^{(t-t_{max})/t_{rise}},
\end{equation}
to the light curve, where $\Delta S$ is the amplitude of the flare,
$t_{max}$ is the peak location of the flare and $t_{rise}$ is the rise
time of the flare. The decay time of the flare has been frozen to 1.3 times the rise time,
   as in \cite{lahteenmaki99b}.
The fitting is done using the MultiNest Markov-Chain Monte Carlo (MCMC) algorithm
\citep{feroz08,feroz09,feroz13}. As shown
in Fig.~\ref{fig:cc1} middle panel, an exponential function fits the 15\,GHz flare
fairly well. From the fit we obtain the rise time of the flare
 $t_{rise}=10.6\pm0.5$~d and amplitude $\Delta S = 0.52\pm0.01$~Jy,
where the uncertainties are estimated from the MCMC analysis.

Assuming an emission region with the geometry of a uniform disk, we
can estimate the lower limit of the variability brightness temperature
\citep{lahteenmaki99,hovatta09}
\begin{equation}\label{eq:tb}
T_{\rm var} = 1.548 \times 10^{-32} \frac{\Delta S d_L^2}{\nu^2t_{rise}^2(1+z)},
\end{equation}
where $d_L$ is the luminosity distance to the object in meters  (here, $4.0\times10^{24}~{\rm m}$),
$z$ is the redshift, $\nu$ is the
frequency in GHz, $t_{rise}$ is in days, and $\Delta S$ is in janskys. This gives us $T_{\rm var} =
(5.2\pm0.5)\times10^{12}$K.

 The variability brightness temperature is related to the Doppler
factor, $\delta$, as $\delta = (T_{\rm var}/T_{\rm b,int})^{1/3}$, where $T_{\rm
  b,int}$ is the intrinsic brightness temperature
\citep[e.g.,][]{lahteenmaki99}. As the variability brightness
temperature estimate is a lower limit, the Doppler factor estimates
are also lower limits. The largest uncertainty in the Doppler factor
estimate comes from the uncertainty in the intrinsic brightness
temperature and what method is used to estimate it.  Assuming
equipartition between the particles and magnetic field, $T_{\rm b,int}
= 10^{11}$K \citep{readhead94}. This results in a Doppler factor
$\delta > 3.7$. If we use a value of $T_{\rm b,int} =
5\times10^{10}$~K, as determined by \cite{lahteenmaki99}, the Doppler
factor is $\delta > 4.7$.

We can further constrain the intrinsic brightness temperature by
comparing the variability brightness temperature with the brightness
temperature obtained via simultaneous Very Long Baseline Array (VLBA)
observations \citep{lahteenmaki99, hovatta13}. In order to study the
parsec-scale jet structure after the 2012 flare we conducted 5 epochs
of target-of-opportunity VLBA observations of Mrk~421 at several
frequency bands (J. Richards et al. in preparation). The first one of
these was taken on 12 October 2012 when the radio flare was
already decaying. The brightness temperature estimate from the 15\,GHz
data is $T_{\rm VLBI} = 5.2\times10^{10}$~K (for a uniform disk), which depends on the
Doppler factor as $\delta = T_{\rm VLBI} / T_{\rm b,int}$. We can then
solve for the intrinsic brightness temperature by calculating $T_{\rm
  b,int} = \sqrt{T_{\rm VLBI}^3/T_{\rm var}} = 5.2\times10^{9}$~K. If
we use this value for the intrinsic brightness temperature and the
variability brightness temperature, we obtain $\delta > 10$.
Thus, we conclude that the lower limit of the Doppler factor is
$\delta \sim 3-10$.

We note that the intrinsic brightness temperature obtained using
  the VLBA data is about $10-20$ times below the equipartition
  limit. We think the peak brightness temperature is likely higher than our estimate because
by the time of the first VLBA epoch,
  the single-dish flux density had already declined by 30 per cent
  from the peak. Therefore it is likely that the true simultaneous
  brightness temperature from the VLBA observations is at least 30
  per cent higher, because the core could have also been more compact,
  increasing the brightness temperature even further. Because of the
  strong dependence of $T_{\rm b,int}$ on $T_{\rm VLBI}$, any
  uncertainties in the VLBA parameters are magnified in the estimate
  of $T_{\rm b,int}$. A slightly higher $T_{\rm VLBI}$ would also
  agree with estimates from \cite{lico12}, who found the core
  brightness temperature of
  Mrk~421 to be of the order of few times $10^{11}$~K, in agreement
  with equipartition arguments.

\subsection{The flare of 2013}\label{sect:2013}
In April 2013, Mrk~421 was again flaring in the  X-ray to TeV bands \citep{balokovic13,paneque13},
reaching the highest levels ever observed at TeV energies \citep{cortina13}.
Triggered by this activity, we began monitoring the source more
frequently at CARMA.

The appearance of this flare was very different from the 2012
flare, both in the $\gamma$ rays and radio. In $\gamma$ rays the
  activity began in March 2013 reaching the highest peak on
  14 April 2013 (MJD 56397).
The flux
increased from about $2\times10^{-7}$\,ph\,cm$^{-2}$\,s$^{-1}$ to
($6.7 \pm 0.9)\times10^{-7}$\,ph\,cm$^{-2}$\,s$^{-1}$ over about  30~d and the flaring
period consisted of several peaks over a total duration of about  100~d. At 15\,GHz, the flux density began increasing in April 2013 from
about 0.6~Jy to its peak of 0.77~Jy on 18 June 2013 (MJD
56461). Similarly, at 95\,GHz the flux density increased from about
0.5~Jy to a peak of 0.71~Jy on 8 June 2013 (MJD 56451).

\begin{figure*}
\includegraphics[scale=0.6]{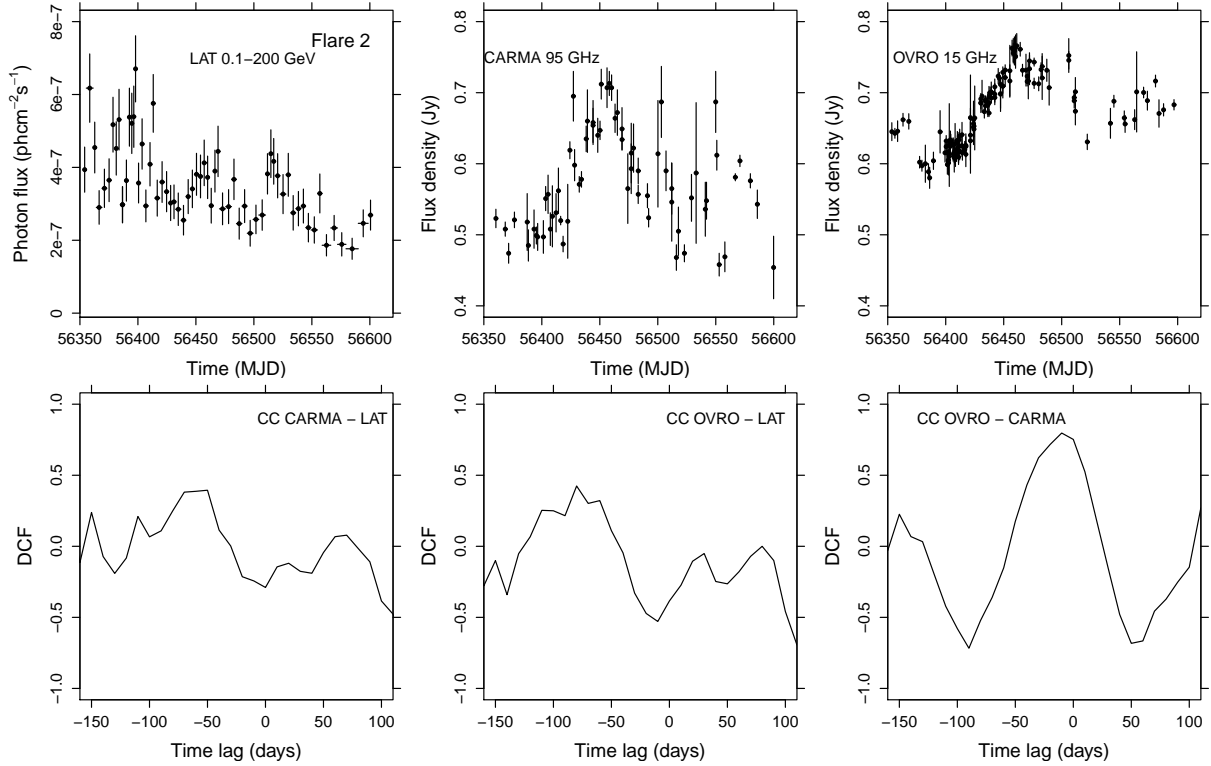}
\caption{Top: Light curves at $\gamma$ rays (left), 95\,GHz (middle),
  and 15\,GHz (right) during the 2013 flare. Bottom: Discrete correlation functions between the
  light curves. Negative time delay means that higher frequency
  emission ($\gamma$-ray or 95\,GHz) is leading. The time range
  corresponds to the dashed lines in Fig.~\ref{fig:lc}.}\label{fig:cc2}
\end{figure*}

The DCFs between all the bands are shown in
Fig.~\ref{fig:cc2}. The highest correlation is found for the 15
and 95\,GHz data, with a peak at a delay of about $-10$ to  $-20$~d, with 95\,GHz leading. The time delays between the
  CARMA (OVRO) and LAT data can only be estimated at about  $-60$~d
  ($-70$~d) because of the broad peaks of the DCF. These estimates
  are, however, compatible with the  broad peak obtained from  our
  cross-correlation analysis of the full light curves,  and thus we
  assume that the events are physically connected in the radio and
  $\gamma$-ray bands.

\section{Conditions in the flaring region}\label{sect:model}
 In this section we attempt to explain the rough features of the 2012
 and 2013  flaring periods, e.g., time delays between various energy bands, peak fluxes and pulse profiles,
by adopting the simplest possible theoretical framework (the single zone
SSC model), together with a minimum set of
free parameters.
In particular, we restrict our modeling  to the
  major flares observed in the periods of MJD 56060-56225 and MJD
  56360-56610, which are denoted  as ``flare 1'' and ``flare 2''' 
  in Sect.~\ref{sect:lc} (see Fig.~\ref{fig:lc}), while modeling 
  of the smaller amplitude variability seen in both bands lies out of the scope of this work.
As the `goodness' of the fits was not the main point of interest here, we did not attempt
a detailed parameter space search for finding the set with the best
$\chi^2$ value.

Based on the long-term radio light curve shown in Fig.~\ref{fig:lc},
the 2013 radio  and millimeter-band flares are more typical of blazar radio emission than the 2012 radio  flare:
they are less sharp, have a decay timescale larger than their rise timescale,  and
the flux density increases by no more than a factor of $\sim 1.4$ in
both  the radio and millimeter bands. We start our analysis with the 2013 flaring events in the context of a typical SSC
model. Then, we highlight the differences between the 2012 and 2013 radio flares and discuss possible modifications within
the same framework that may explain the 2012 data. 
  Our main goal  is to 
investigate what conditions are required to produce this extreme and unique radio
flare. 
 As already discussed in Sect.~\ref{sect:lc}, we assume that the events in the
radio and $\gamma$-ray bands are physically connected.
We note that, in what follows, we ignore the small differences between quantities measured in the observer frame and the source frame
since $1+z \simeq 1$.

\subsection{The $\gamma$-ray and radio flares of 2013}
\begin{figure}
\centering
  \includegraphics[width=0.45\textwidth]{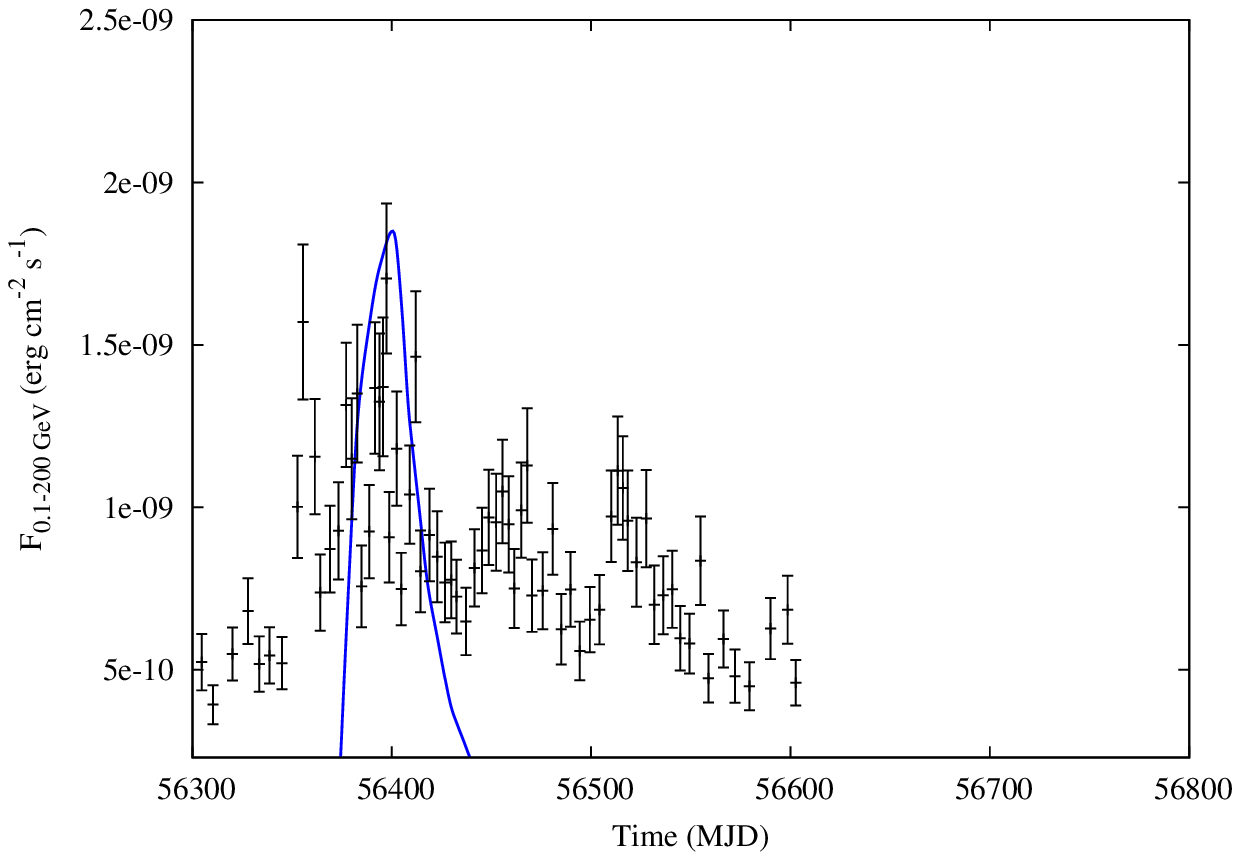} 
  \includegraphics[width=0.45\textwidth]{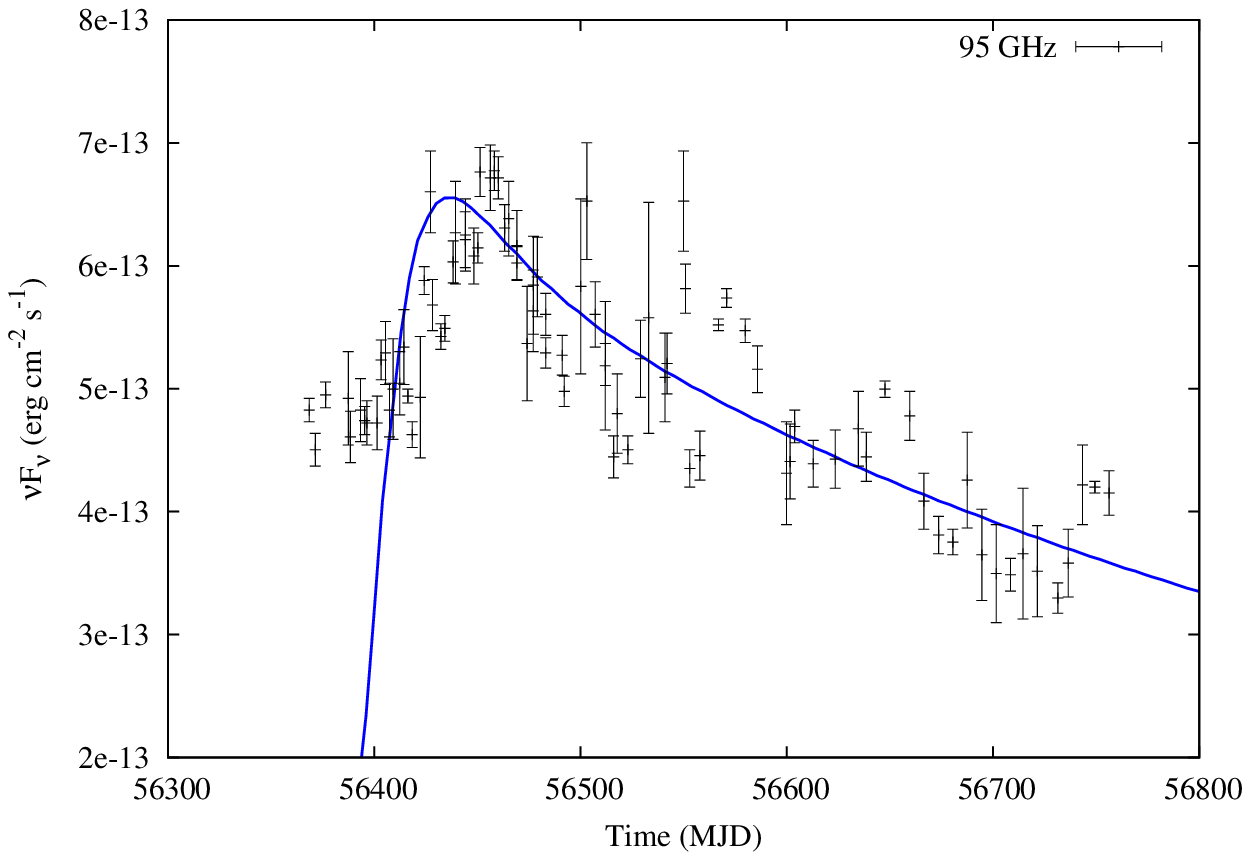}
  \includegraphics[width=0.45\textwidth]{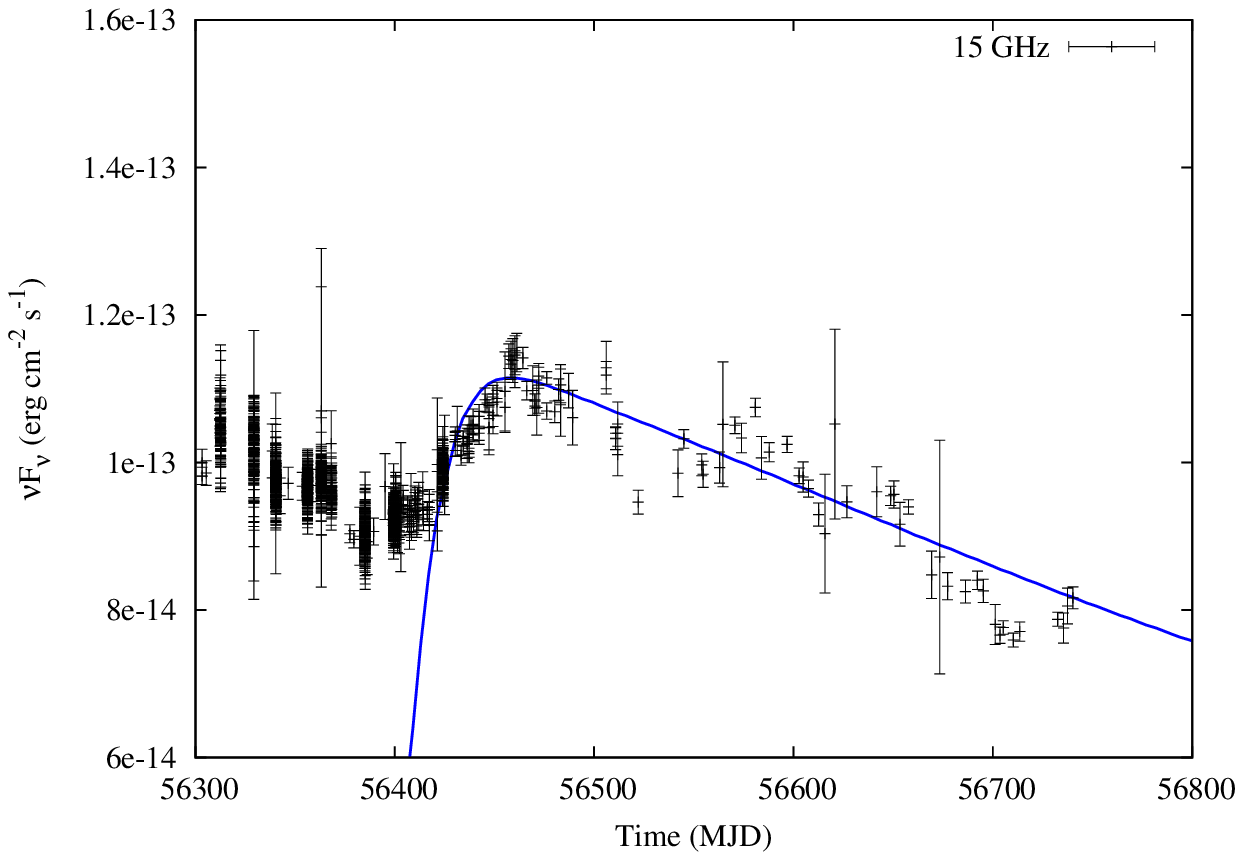}
\caption{{\sl Fermi}-LAT light curve  (top), CARMA 95~GHz light
  curve  (middle) and OVRO 15~GHz light curve  (bottom).
The results of the one-zone SSC model described in the text are shown
with  blue lines in all panels. The time range of the fit corresponds
to the 2013 flare indicated by the dashed lines in Fig.~\ref{fig:lc}.}
\label{lc}
\end{figure}
Motivated by the fact that the CARMA and OVRO fluxes peak $\sim 60$
and  70~d after the $\gamma$-ray flare, respectively
(Sect.~\ref{sect:2013}), and by the fact that the decaying part of
their light curve is approximately exponential, we applied the
simple scenario where both the $\gamma$-ray and radio flares originate
 in the same region, which does not necessarily have the
same physical conditions as the region responsible for the quiescent
emission. We note that  we do not attempt to model the
  sub-structure of the flares, but we only consider the highest
   peak in each band.  In particular, we focus on
  the largest $\gamma$-ray  flare, which peaks at MJD
  $\sim$56400. We also note that the DCF only shows a
  single peak for this  flare, unlike in 2012 where we will
    consider multiple associations between flares.

 In our  scenario  the $\gamma$-ray flare is produced
by an instantaneous injection event of electrons having Lorentz factor
$\gamma_0$ and emitting in $\gamma$ rays through Compton scattering of
synchrotron photons (SSC).  In particular, electrons are injected at
$\tau=0$, which is set to be the time of the $\gamma$-ray flare, and
then they are left to evolve  via synchrotron and SSC
cooling.  In this context, the observed delays $t_{\rm h} \sim 60$~d
and $t_{\ell} \sim 70$~d between the $\gamma$-ray and the radio
flares at $\nu_{\rm h}=95$~GHz and $\nu_{\ell}=15$~GHz, respectively,
correspond to the cooling timescale of electrons that have been
injected at  $\tau=0$ with Lorentz factor $\gamma_0$.

\subsubsection{Analytical estimates}
To define the size $R$ of the emitting region we choose
as a typical variability timescale ($t_{\rm v}$) the one dictated by
the $\gamma$-ray light curve. 
 For the purpose of our analysis,  we
choose a variability timescale of $t_{\rm v}=20$~d based on the light
curve. We note that the exact value is not critical as the estimate
will be refined in the next section where we model the flares numerically.
We find that $R \approx  2 \times 10^{17} \ {\rm cm}\ \left(\delta/4 \right) \left(t_{\rm v}/ 20 \ {\rm d}\right)$,
where we normalized the Doppler factor to 4 (see Sect.~\ref{sect:2012}).
In this framework, one can derive the magnetic field strength $B$ of the emission region as a function
of the radio frequency $\nu_{\rm h}$, the observed time delay $t_{\rm h}$ and the Doppler factor $\delta$.
Electrons that have been injected at  $\tau=0$ with Lorentz factor $\gamma_0$ cool due to synchrotron losses
and reach a Lorentz factor $\gamma_{\rm h}$. This can be found by solving the characteristic equation for synchrotron cooling
and is given by
\eqb
\gamma_{\rm h} = \frac{1}{1/\gamma_0+ \alpha \delta t_{\rm h}} \approx \left(\alpha \delta t_{\rm h} \right)^{-1},
\label{gamma}
\eqe
where $\alpha = (4/3)\sth c  U_{\rm B}/ m_{\rm e} c^2$ and
  $U_{\rm B}=B^2/8\pi$ is the energy density of the magnetic field. The
approximation holds as long as $\gamma_0 \gg \gamma_{\rm h}$.
Combining Eq.~(\ref{gamma}) with the characteristic synchrotron
frequency $\nu_{\rm h}=\delta m_{\rm e} c^2 b \gamma_{\rm h}^2/ h$,
where $b=B/B_{\rm cr}$, $B_{\rm cr}=4.4 \times 10^{13}$~G, and
  $h=6.63\times10^{-27}$~erg s is the Planck constant, we find that the required magnetic field strength of the region is
\eqb
B \simeq 0.5  \ {\rm G} \left(\frac{\nu_{\rm h}}{95 \ {\rm GHz}}\right)^{-1/3} \left(\frac{\delta} {4}\right)^{-1/3} 
\left(\frac{t_{\rm h}}{60\ {\rm d}}\right)^{-2/3},
\label{B-delay}
\eqe
 which  depends most strongly on the time delay between the $\gamma$-ray and
radio flares. Notice also that lower values of the Doppler factor  favour stronger magnetic fields. The expected time delays for the two frequencies should satisfy the relation
$t_{\ell}/t_{\rm h}=(\nu_{\rm h}/ \nu_{\ell})^{1/2}$.
This is roughly consistent with the observed values, since
$\sqrt{\nu_{\rm h}/ \nu_{\ell}}=2.5$ and  $t_{\ell}/t_{\rm h}$ is in
the range of $1 - 1.7$ based on the broad peaks in the time
delays (see Fig.~\ref{fig:cc2}). 
 The duration of a flare is also related to the magnetic field strength, since it depends on the electron cooling timescale.
For a given frequency $\nu_{\rm h} \propto B \gamma^2_{\rm h}$ the full width at half maximum (FWHM) of a flare
will scale roughly as $\propto B^{-2} \gamma_{\rm h}^{-1} \propto B^{-3/2} \nu_{\rm h}^{-1/2}$. 
Thus, for strong enough magnetic fields the duration
of a flare even  at low frequencies may be shortened significantly.

Assuming that the inverse Compton scattering occurs in the Thomson
limit, which will be checked {\sl a posteriori}, the typical
  energy of up-scattered photons is given by $E_{\gamma} \simeq
  (4/3)\delta m_{\rm e} c^2 b \gamma_0^4$. Using the value of $B$ derived above,
we find that the injection Lorentz factor of electrons is
$\gamma_0 \simeq 2\times 10^{4} \left(E_{\gamma}/ 4 \ {\rm GeV}\right)^{1/4} \left(\delta/4 \right)^{-1/4} \left(B / 0.5 \ {\rm G}\right)^{-1/4}$,
where we chose $E_{\gamma}=4$~GeV as a representative value for the
energy of $\gamma$-ray photons.
Because of the weak dependence of
$\gamma_0$ on $E_{\gamma}$, a different choice of the $\gamma$-ray energy would
not affect the derived value for the injection Lorentz factor.   Increasing $E_{\gamma}$ by a factor of 50, for example, would increase $\gamma_0$ by only a factor of $\sim 2$.
In principle, electrons could have been injected with $\gamma > \gamma_0$
and still produce the GeV flare, since they would very quickly cool to the
value $\gamma_0$. In this case a
contemporaneous TeV flare would  also be expected, and indeed one was seen by
the MAGIC and VERITAS instruments \citep{cortina13}.
Having derived the expression for $\gamma_0$, which does not depend strongly on the parameters,
we can now verify  our earlier assumption  that the Thomson limit  applies. Since $b \gamma_0^3 = 0.09 \left(\frac{B}{0.5 \ {\rm G}}\right)^{1/4}
\left(\frac{E_{\gamma}}{4 \ {\rm GeV}} \right)^{3/4} \left(\frac{\delta}{4} \right)^{-3/4}   < 3/4$, we are safely in the Thomson regime.
The derived values for the magnetic field strength and the Lorentz factor of injected
electrons are reasonable and consistent with typical SSC models of
 high frequency peaked BL~Lac objects, such as Mrk~421 \citep[e.g.,][]{abdo11,aleksic12}. Note, however, that
here we adopted a low value for the Doppler factor implied by radio observations, in contrast to typical
SED modeling  where much larger values, e.g. $20-50$, are usually used
\citep[see e.g.,][]{maraschi99}

\subsubsection{Numerical results}
The required electron injection luminosity is determined numerically
by fitting the observed energy fluxes. For this, we employed
the numerical code described in \cite{mastichiadis95, mastichiadis97}. Using as a stepping stone
the values determined analytically in  Sect. 4.1.1., we  derive the following set of parameters:
$R=10^{17}$~cm, $B=0.1$~G, $\delta=2.2$, $\gamma_0=2\times 10^4$ and  the electron injection compactness $\ell_{\rm e}^{\rm inj}=1.6\times 10^{-2}$. This is
defined as $\ell_{\rm e}^{\rm inj} = \sth R U_{\rm e}/ m_{\rm e}c^2$  where  $U_{\rm e}$  is the energy density of electrons
at injection  time as measured in the rest frame of the emission region and $\sigma_T=6.65\times10^{-25}$~cm$^2$ is the Thomson cross section.
Although our analytical estimates were based upon the hypothesis
of instantaneous injection, here we find that an injection episode lasting $\sim R/c$, or equivalently $20$~d,
is better in reproducing the observed flares.

Our results are  illustrated in Fig.~\ref{lc}, where we
  use energy fluxes to facilitate comparison between the different
  bands.  In the radio band,  flux densities are converted to
    energy fluxes by multiplying by the observing frequency and
  converting to cgs units. The energy  fluxes in the
   $\gamma$-ray band  are obtained  as
    described in Sect. 2.   The model light curves can describe
  the radio observations fairly well, although the  model
  $\gamma$-ray flare is slightly broader than the actual
  data.  Still, the results of Fig.~\ref{lc} are
  satisfactory given the small number of free parameters and the fact
  that we have not attempted to find the best fit (i.e., with the
  lowest $\chi^2$ value). The radio flares in both frequencies have
  wide pulse profiles and exhibit a slow exponential decay after their
  peak. Both of these  features contrast with the sharpness
  and the symmetry of the 2012 flare.  In   the
    present scenario, the width of each flare is related to the
cooling timescale of electrons emitting at the particular energy band
and, thus, the radio flares are wider than the one in $\gamma$ rays
 (see Sect. 4.1.1.). The asymmetry of the pulse
 profiles at lower energies is a strong prediction of this
model.  Possible expansion of the  flaring region and/or
decay of the magnetic field would  increase the predicted
  asymmetry.

For the derived parameters, the region is initially\footnote{The
  electron energy density will actually decrease with time because of
  (i) cooling and (ii) no replenishment of particles in the region.}
particle dominated with $U_{\rm e}\approx0.18$~erg cm$^{-3}$ and
$U_{\rm B}=4\times 10^{-4}$~erg cm$^{-3}$.  The value of the Doppler
factor derived here is approximately half of that obtained in
Sect.~\ref{sect:2012} assuming equipartition. However, the large ratio $U_{\rm e}/U_{\rm B} \simeq
450$  derived from the values mentioned above would
also imply a higher brightness temperature \citep[e.g.,][]{readhead94}
by a factor of $450^{1/8}\sim 2$ and thus a lower value of $\delta$ by
a factor of $2^{-1/3} \sim 0.7$. Given that the fit shown in Fig.~\ref{lc} is not unique, one could
find parameters that would decrease the initial ratio $U_{\rm
  e}/U_{\rm B}$ and bring the emission region closer to an
equipartition state.  By assuming that a fraction $\eta_{\rm rad}$ of
the injected luminosity in electrons is radiated as $\gamma$ rays,
i.e. $\eta_{\rm rad} \ell_{\rm e}^{\rm inj} = L_{\gamma}^{\rm obs}
\sigma_{\rm T}/4 \pi R \delta^4 m_{\rm e}c^3$, and using the
definition of $\ell_{\rm e}^{\rm inj}$ as well as the equations for
$R$ and $B$ in Sect.~4.1.1., we find that 
\eqb \frac{U_{\rm e}}{U_{\rm
    B}} \simeq F_0
    \left(\frac{\eta_{\rm
      rad}}{0.1} \right)^{-1}\!\! \left(\frac{\delta}{4}
\right)^{-16/3} \!\!  \left( \frac{\nu_{\rm h}}{95 \ {\rm
      GHz}}\right)^{2/3} \left(\frac{t_{\rm h}}{60 \ {\rm d}}
\right)^{4/3} \left( \frac{t_{\rm v}}{20 \ {\rm d}}\right)^{-2}, \eqe
where $F_0 \equiv F_{\gamma}^{\rm obs} / 2\times 10^{-9}$~erg
cm$^{-2}$ s$^{-1}$. The ratio of energy densities  depends
  strongly on the Doppler factor. Thus, searching for possible fits
with a slightly higher value of the Doppler factor, e.g., $\delta
\simeq 7$, would bring down the ratio from 450 to close to unity and
would ensure rough equipartition between the particles and magnetic
field at injection.

\subsection{The $\gamma$-ray and radio flares of 2012}
  \begin{figure*}
\centering
\includegraphics[width=0.45\textwidth]{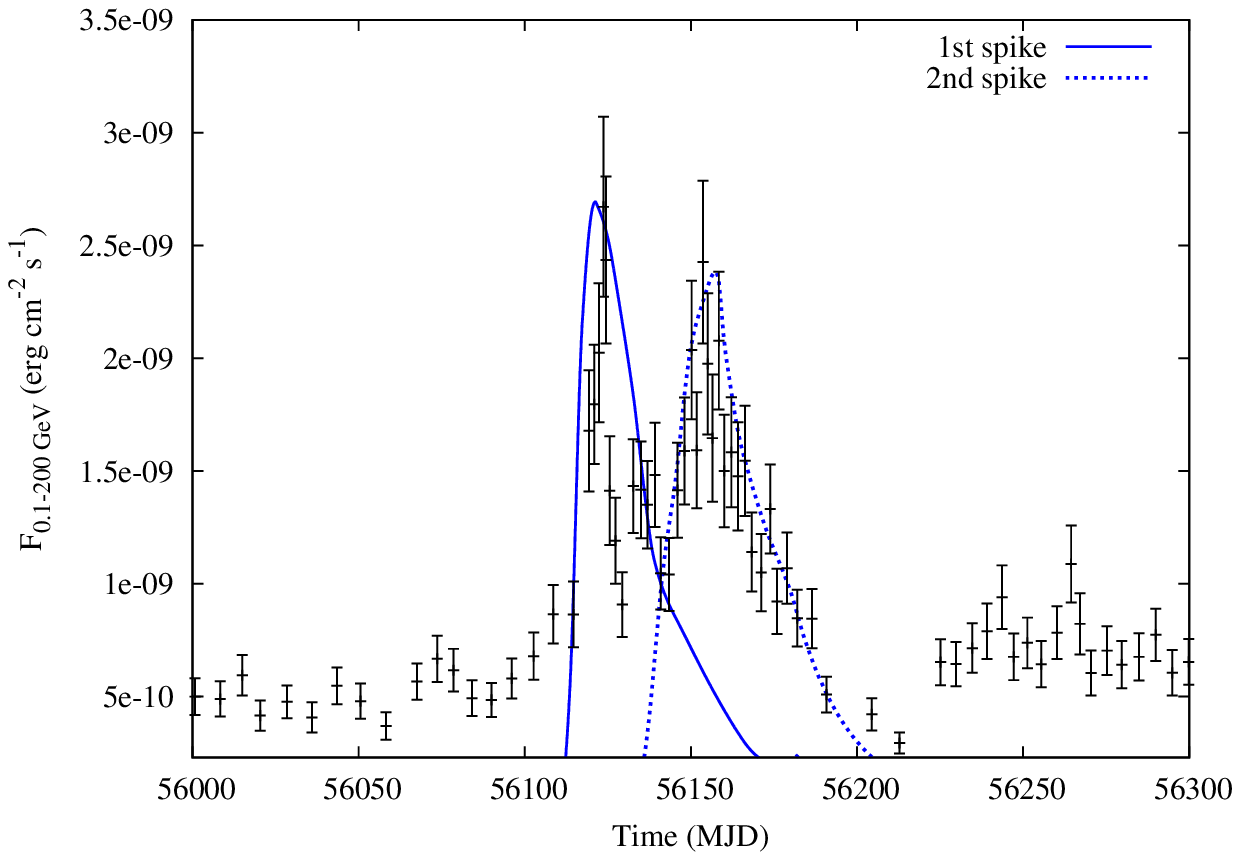}
\includegraphics[width=0.45\textwidth]{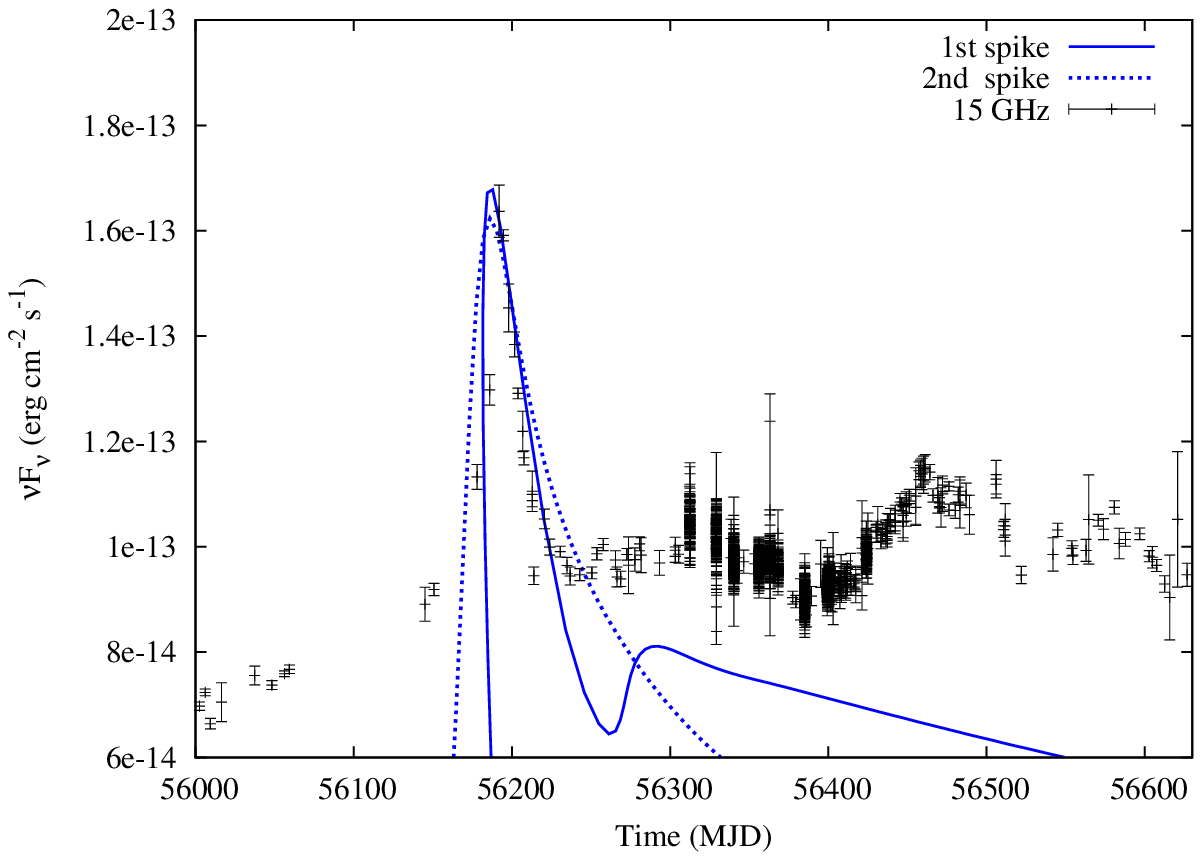}
\caption{{\sl Fermi}-LAT light curve (left) and OVRO 15~GHz light
  curve (right)  in energy flux units along with the light curves (blue lines) obtained
from the model.  Solid and dashed lines 
demonstrate the model results, if the radio flare is associated with the first and second spikes of the $\gamma$-ray
flare, respectively. A different x-axis range is used in the right panel to show the extended model light curve more clearly. 
}
\label{2012}
\end{figure*}
The 2012 radio flare is a unique event not only because of its large
flux increase but also because of its symmetric pulse profile, which
resembles flares at higher energies where the  corresponding
radiating particles cool efficiently.   According to the
  cross-correlation analysis presented in Sect. 3.1, the 2012 radio
  flare may be associated with either the first or second spike of the
  $\gamma$-ray flare (see Figs.~1 and 2).  We investigate both
  possibilities under the assumption that there is indeed a physical
  connection between the extreme radio and $\gamma$-ray flares. For
  this we  begin with the parameters we derived for the 2013
  ``typical'' flare, and examine two alternative scenarios. These
  include changes of the Doppler factor and magnetic field strength,
  since both parameters are related to the time delay (see
  Eq.~\ref{B-delay}) and to the pulse profile (see Sect. 4.1.1.).  We
  emphasize that in both scenarios we attempt to introduce minimal
  changes to the parameters.  As before, we do not consider
  the  small-amplitude sub-structure of the flares. 
  
\subsubsection{Association of the  radio flare with the first
    $\gamma$-ray spike}   We first investigate the
  possibility that the radio flare is associated with the
   narrower first spike of the $\gamma$-ray flare,  which peaks at
  $\sim$56133 MJD.  In order to do this, we
 use the same parameter set as  derived for the 2013 flare, but
 add variations in the Doppler beaming  to the
  model. Such changes could occur if the emission region moves on a
curved trajectory with a changing viewing angle, or if the bulk
Lorentz factor of the emission region changes. Varying Doppler factors
have been previously used to explain radio spectral variations in, for
example, S5~0716+714 \citep{rani13}.  The Doppler factor $\delta$ is
modeled as \eqb \delta(\tau) =\delta_0 \left(1+ g (\tau)
  \cos(2\pi \tau / P) \right), \eqe where $\delta_0=2.1$, $\tau$ is
the time in the comoving frame in $R/c$ units\footnote{Time is
  measured with respect to the time of the first spike seen in the
  2012 $\gamma$-ray flare.}, $P=5.4$ is the period of the
variation in $R/c$ units and $g (\tau)$ is the function 
\eqb
g (\tau) & = & 0.3, \ {\rm for} \ \tau < \tau_{\rm br} \\
g (\tau) & = & 0.3 e^{-(\tau-\tau_{\rm br})/T} , \ {\rm for}
\ \tau \ge \tau_{\rm br}, \eqe  with $\tau_{\rm br}=1.2P$
and $T=0.2 P$.  

 The  model-derived pulse profiles are shown in Fig.~\ref{2012} with blue solid lines.
Although the cooling timescale of electrons emitting at radio
frequencies is long compared to those emitting initially at $\gamma$
rays, the shape of the radio flare (right panel in Fig.~\ref{2012}) is
well reproduced here, because of the change in the Doppler factor,
which does not allow the observer to see the typical wide pulse
profile (see e.g., Fig.~\ref{lc}).  The effect of the variable
$\delta$ on the $\gamma$-ray light curve is
 negligible because the radiating electrons at the time of
injection have short cooling timescales.  Note that the exponentially
decaying amplitude in the variation of $\delta$ is required in order
to avoid any excess of the radio emission following the extreme flare.

 Assuming that the change in the Doppler factor is caused by small variations
 in the viewing angle $\theta$ while the Lorentz factor
remains constant at $\Gamma=10$,   no extreme variations
  in either $\delta$ or the angle $\theta = \cos^{-1} \left(\beta^{-1}
    \left(1-(\Gamma \delta)^{-1} \right) \right)$ are
   required.\footnote{This  Lorentz factor is approximately five times
      higher than the value derived in Sect.~\ref{sect:discussion} for
      the bulk Lorentz factor of the jet ($\Gamma_{\rm var}=1.9$). If
      we had used this value instead, only the angle $\theta$ would be
      affected, i.e. it would be larger by a factor of 2.  However, it
      is still possible that the Lorentz factor of the emission region
      is higher than that of the jet \citep[see
      e.g.,][]{nalewajko14}.}
    In particular, $\delta$ and $\theta$
  change by less than $28$ and $22$ per cent, respectively,  relative to their
  average values.

  The fact that the necessary parameter changes are small
  make this scenario plausible and attractive. 
 However,  if we attempt to interpret the second spike of the
$\gamma$-ray flare in the same framework, then we cannot avoid the
appearance of a radio flare  $\sim$~70~d  later,
i.e. at MJD~56220.   Its absence implies one (or more) of
the following: (i) a faster onset of the exponential decay of
$\delta$; (ii) a different functional form for $\delta(\tau)$; (iii)
different conditions in the emission region  which suppress
the 15~GHz emission, such as different magnetic field and/or
size; (iv) a different framework for the second $\gamma$-ray spike.
  
 Finally, the choice of a damped oscillator for modeling $\delta(\tau)$, albeit plausible,
was not based on a physical picture. 
 According to the discussion of Sect. 4.1.1., the FWHM of a flare depends on the magnetic
field strength.   Therefore, another alternative might be to adopt
  a stronger magnetic field for the emission region.
Although a  stronger magnetic field would
lead to a shorter duration of the flare, an even lower value of the Doppler factor would be required in order
to explain the observed time delay between the $\gamma$-ray and radio flares (see Eq.~\ref{B-delay}).
Given that the value of the Doppler factor  already lies at the low end of  values consistent with observations, we do not examine this possibility in more detail.

\subsubsection{Association of the  radio flare with the second $\gamma$-ray spike}
  We now consider the scenario where the radio
  flare is associated with the second spike of the $\gamma$-ray flare.
 This scenario would arise naturally if the first $\gamma$-ray spike occurred
when the emission region was still optically thick  to radio emission.
In this case, the time-delay between the $\gamma$-ray and radio flares is shorter than before, which, in our framework, suggests
faster electron cooling. 
We searched, therefore, for  reasonable fits using a stronger magnetic field than
 that adopted in Sect. 4.2.1, and  obtained the 
following parameter set: $B=0.25$~G, $\delta=2.3$ and $\ell_{\rm e}^{\rm inj}=1.3\times 10^{-2}$.  All other parameters, including the injection profile, are the same as before.

The  resulting pulse profiles are shown in Fig.~\ref{2012}
with dashed blue lines.  We find that the light curves obtained by the
model are in rough agreement with the observations, with the radio
light curve having a slightly longer decay timescale than  is
  observed.   In this scenario, even without Doppler factor
 variations, we  obtain a sharper radio flare than
the one in 2013 (see e.g. Fig.~4).  Comparing these parameter
  values with the first scenario, we see that although we
  adjusted the numerical values for three parameters, only the
  magnetic field is significantly altered. It has increased by a
factor of 2.5.  We can compare the FWHM of the 15~GHz radio flares in
2012 and 2013 as obtained by our model (Figs.~\ref{lc} and
\ref{2012}).  We find their ratio to be $\Delta T_{2012}/\Delta
T_{2013} \sim 0.16$, where $\Delta T$ denotes the duration at
FWHM. This is  comparable to the analytical
estimate we can obtain from Eq.~\ref{B-delay},  $\Delta T_{2012}/\Delta T_{2013} \simeq \left(B_{2012}/B_{2013}\right)^{-3/2} \simeq
\left(0.25/0.1\right)^{-3/2}= 0.25$.

 We note that this model is subject to tight constraints on
  its parameters. These must be adjusted to ensure: (i) the electron
cooling timescale  results in the observed delay between the
$\gamma$-ray spike and the radio flare (see Eq.~(\ref{B-delay})); (ii)
the cooling timescale of electrons radiating at the radio frequency
 is such that it explains the observed width of the pulse; and
(iii) the synchrotron self-absorption frequency, which depends on
$\ell_{\rm e}^{\rm inj}$, $B$ and $R$,  is below the radio
frequency where the extreme flare is observed. From the above it
becomes clear that fine tuning of the parameters is required for this
scenario to be viable.

\section{Discussion}\label{sect:discussion}
Unlike in many other blazars, prominent radio flares are rare in Mrk~421 and there are  few
examples where the radio flares have been modeled in detail. In 1997 a
22\,GHz flare was observed with the Mets\"ahovi Radio Observatory 14-m
telescope \citep{tosti98}. The flare lasted about  200~d during
which the flux density increased by a factor of two. The radio flare occurred
 60~d after a large optical flare that also included very high
optical polarization. Due to the high optical polarization during the
flare, the authors interpreted the flare within the shock-in-jet model
of \cite{marscher85}. The delay between the optical and radio bands was
explained  by opacity effects. While this model may explain
the general features of the flares and the delay between optical and
radio bands, more detailed modeling was not presented by \cite{tosti98}.

In 2001 a simultaneous flare was observed at TeV, X-ray and 5\,GHz
radio bands \citep{katarzynski03}. The radio flux density increased by
a factor of 1.5 and the flare lasted only for  15~d. The absence of
time delays between the bands indicates that the emission region was
optically thin even at radio frequencies.   An instant injection of particles into the
  jet base was unable to explain the radio flares as the
  electrons would have cooled too much to generate significant flux
  density changes before reaching the optically thin regime for radio
  emission. Instead, 
detailed modeling of the
flare suggested that it was caused by  in-situ acceleration of electrons in a
blob in the jet. The expansion of the blob explains the decay of the
flare at all bands. 

The 2013 flaring event was a more typical example of radio flares in
Mrk~421, with a broad flare in the millimeter and centimeter bands
(see Fig.~\ref{fig:lc} for the long-term behavior). Assuming that the
events in the different bands are physically connected, we modeled the
most prominent features of this event with a simple one-zone SSC model
where an instantaneous injection of electrons produces a $\gamma$-ray
flare. Then, as electrons cool because of synchrotron and SSC energy
losses, they radiate at longer wavelengths and eventually produce the
radio flares with a delay relative to the $\gamma$-ray flare. Within
this context, we estimated the required magnetic field strength, the
size of the emission region, and the Lorentz factor of electrons at
injection.  While the simple model presented here can describe the
main features of the data satisfactorily, more information from
simultaneous multi-wavelength observations is needed in order to
discriminate between other possible models. For example, one could
postulate a scenario where the delay between the $\gamma$-ray and
radio flares is related to opacity effects, such as if the source were
optically thick in radio frequencies at the time of the GeV flare but
eventually became optically thin due to the expansion of the emission
region or decay of the magnetic field \citep[e.g.,][]{fuhrmann14}.
Although the  coincident unprecedented events between the LAT
and OVRO data suggests a physical connection of the high- and
low-energy emission, models where two spatially separated emission
regions are invoked \citep{blazejowski05, petropoulou14} are still
viable alternatives.  We note that the 2013 flare occurred during a
planned multi-wavelength campaign and Mrk~421 was observed regularly
with numerous instruments at various frequency bands. Therefore we
anticipate several dedicated studies of the behavior of Mrk~421 during
this flare.

Even though the 2012 radio flare was  more extreme, the
Doppler factor inferred from the radio variability time-scale is
fairly low, only about $3-10$. This is in accordance with the low observed
apparent speeds obtained through radio interferometric observations
where only subluminal speeds
have been detected \citep[e.g.,][]{piner99,lico12, lister13}. Based on
the jet brightness asymmetry, low observed apparent speeds, and low
brightness temperature of the core component, \cite{lico12} estimate
the radio Doppler beaming factor to be about 3, consistent with our
estimate.  If simultaneous X-ray or TeV data were available, we could
estimate a minimum Doppler factor by demanding the source to be optically thin to 
the highest energy emitted photon 
\citep[e.g.,][]{dondi95}. From previous TeV flares of Mrk~421 it has
been estimated that the minimum Doppler factor required for the TeV
emission to be optically thin is $\sim 10$
\citep{celotti98}, which is marginally higher than the value obtained
from our radio data when equipartition is assumed.

Preliminary analysis of our follow-up VLBA observations indicate that
the component speeds were consistent with subluminal motion even after
the major flare \citep{richards13}. If we take the fastest observed
jet  speed, $\beta_\mathrm{app}=0.28c$, obtained
 from several years of monitoring  at 15\,GHz with
the VLBA~\citep{lister13}, we can estimate the jet Lorentz factor  as
\begin{equation}
\Gamma_\mathrm{var} = \frac{\beta_\mathrm{app}^2 +
  \delta_\mathrm{var}^2 + 1}{2\delta_\mathrm{var}},
\end{equation}
where $\delta_\mathrm{var}$ is the Doppler factor inferred from the
variability time-scales. This results in a Lorentz factor of
$\Gamma_\mathrm{var} \sim 1.7-5$. Similarly, we can estimate the viewing
angle of the jet  to be
\begin{equation}
\theta_\mathrm{var}=
\arctan\frac{2\beta_\mathrm{app}}{\beta_\mathrm{app}^2 +
  \delta_\mathrm{var}^2 - 1},
\end{equation}
resulting in $\theta_\mathrm{var} \sim 0.3-4^\circ$. As
  noted by \cite{lico12}, it is unlikely that the component speeds
  resemble the flow speed in Mrk~421 because unreasonably small
  viewing angles would be required to explain the beaming estimates 
  from the jet/counter-jet ratio. In this case the above equations
  would not be valid for estimating the true Lorentz factor and
  viewing angles. However, they find a viable scenario for
  Mrk~421 with a structured jet, where the viewing angle is between
  $2^\circ$ and $5^\circ$, and the Lorentz factor of the radio
  emitting region about 1.8, in agreement with our estimates.

One of the long-standing problems in modeling of the high synchrotron
peaked blazars is the large discrepancy
between the Doppler factor values inferred from radio observations and those
obtained by SED modeling, with the former being usually $\lesssim 10$
\citep[e.g.,][]{piner99,lister13} and the latter 
lying in the range $\sim 20-50$ \citep[e.g.,][]{maraschi99,abdo11}.  
Several alternatives have been explored to explain
this well-known ``Doppler factor crisis'', such as a structured jet
\citep{ghisellini05}, a decelerating jet \citep{georganopoulos03},
and a jet-in-jet model \citep{giannios09}. In the structured jet model
the high-energy emission comes from a fast spine of the jet while the
radio emission is produced in a slower sheath
\citep{ghisellini05}. This model is favored by observations of limb
brightening of the Mrk~421 jet at 43\,GHz \citep{piner10}.

In the present work, where we have adopted a single-zone emission model,
we attempted to avoid a large discrepancy between the Doppler factor values
inferred from the observations and those 
used in the modeling of the flares. 
However, one could relax this condition and search for possible fits to both the radio and $\gamma$-ray flares  
with $\delta \gtrsim 20$. One can estimate the effect of a higher $\delta$ on 
our results by inspection of the analytical expressions derived in
Sect. 4.1.1. In particular, 
we find that $R\propto \delta$, $B\propto \delta^{-1/3}$, $\gamma_0 \propto \delta^{-1/6}$, $U_{\rm e}/ U_{\rm B}\propto \delta^{-16/3}$
and FWHM$\propto \delta^{1/2}$, where we assumed that the radio observing frequencies and their time-delays
with respect to the $\gamma$-ray flare are fixed. On the one hand, choice of a higher $\delta$ would require
a weaker magnetic field, a larger emission region and would reduce the ratio of particle
to magnetic energy densities (see discussion in Sect. 4.1.2).
On the other hand, the model light curves would be wider than those presented in Fig.~4, since the FWHM would increase. This is a direct
result of the longer electron cooling timescale due to the weaker magnetic field.
We conclude, therefore, that values $\gtrsim 20$ are less plausible for the modeling of both the radio and $\gamma$-ray flares.
Another possibility is to assume a high Doppler factor value for the $\gamma$-ray flare alone and 
a low value for the radio flares.
Because a single-zone model only contains a single Doppler factor,
this parameter would need to change rapidly between the two flares.
This scenario would require the Doppler factor to drop by $\gtrsim
10$ between the $\gamma$-ray and radio flare within a period
of $t_{\rm h}=60$~d, which is much larger than the modest variations
applied in the modeling of the 2012 flares.

 The 2012 $\gamma$-ray flare had two prominent spikes, and we used
  these to investigate the physical conditions required to produce the
  extreme radio event. By using the 2013 model parameters as a
  starting point and introducing as few modifications as possible, we
   considered two possible scenarios. These  result
    from the association of the radio flare with the first or second
  $\gamma$-ray spikes.  Under the assumption of a physical connection
  between the radio flare and the first $\gamma$-ray spike, we showed
  that  the addition of a varying Doppler beaming factor
   to a one-zone SSC model with fairly typical parameters can
    explain the observed sharp radio flare. The SSC model would
    otherwise produce too broad a radio flare, similar to the 2013
   event (see Figs.~\ref{lc}-\ref{2012}).  Although this
  scenario succeeds in explaining the radio flare, it results in a
  wider $\gamma$-ray pulse profile than  is observed (solid
  lines in Fig.~\ref{2012}). Moreover, it  is difficult to
  explain the lack of a second peak in the radio light curve, as
  discussed in the end of Sect.~4.2.1.  Nonetheless, a
    varying Doppler beaming factor is not unreasonable. An adequate
    variation could result from, e.g., a modest change in the Lorentz
    factor or the viewing angle. In the latter case, the sharp flare
  would occur when the  direction of motion of the emission
  region  crosses very near to the line of sight to the observer. Similar models with
  curved emission region trajectories have been suggested for other
  blazars as well
  \citep[e.g.,][]{marscher08,marscher10,abdo10,rani13,aleksic14,molina14}.
  These models assume that either the emission feature moves on a
  helical trajectory due to an ordered helical magnetic field
  \citep{marscher08,marscher10,molina14}, or that there is an actual
  bend in the jet \citep{abdo10,aleksic14}.

  In our second scenario, we considered the possibility
  that the radio flare is associated with the  second,
    broader $\gamma$-ray spike. In this case, the short duration of
  the radio flare  and the shorter time delay between the
  radio and $\gamma$-ray flares imply faster electron cooling, and
   thus a stronger magnetic field than the one used in the
  first scenario. We found indeed that a viable model is achieved by
  increasing the magnetic field strength by a factor of 2.5.
  Fig.~\ref{2012} shows that the model light curve (dashed lines)
  describes well the $\gamma$-ray data, although the
   modeled radio flare is  now not quite as sharp as the observed one.
  In this scenario, the first $\gamma$-ray spike would have to be
  unassociated with the radio event, perhaps by occurring closer to
  the black hole where the emission region is optically thick for
  radio emission. 

\section{Conclusions}\label{sect:conclusions}
We have obtained radio, millimeter and $\gamma$-ray light curves of
Mrk~421 during the flaring activity in 2012 and 2013.  In July 2012,
Mrk~421 exhibited the largest $\gamma$-ray flare observed by {\it
  Fermi} since the beginning of its mission. About
 $40-70$~d later, the largest ever 15\,GHz flare was
observed. The flare rise time determined from an exponential fit was
just  $10.6\pm0.5$~d, which is extreme compared  to
  previous radio flares observed in the source. This
 implies a variability Doppler factor
 $\delta\sim3-10$. In 2013 Mrk~421 underwent major
$\gamma$-ray flaring again, followed by radio and millimeter flares
about  60~d later.   Under the assumption that the
events in the different bands are physically connected  we
  have modeled the variations with the simplest possible theoretical
model, with the main goal of explaining the extreme radio flare in
2012.  Starting with the less extreme 2013 flare, we
 obtained a one-zone SSC model that could explain the main
features of the flares reasonably well.
 The 2012 $\gamma$-ray flare was double peaked, and by
   modeling the radio flare as connected to each peak in
    turn, we were able to reproduce the radio flare with either a
  varying Doppler factor, or an increased magnetic field strength. In
  both cases  several specific conditions in the jet need to be
    fulfilled for the models to be viable, showing the extreme and unique
  nature of the 2012 event.

\section*{Acknowledgments}
We thank Bindu Rani, Jeremy Perkins, Dave Thompson, and the anonymous
referee for their comments and suggestions that greatly improved the paper.
T.\,H. was supported by the Academy of Finland project number 267324.
M.\,P.: Support for this work was provided by NASA
through Einstein Postdoctoral
Fellowship grant number PF3 140113 awarded by the Chandra X-ray
Center, which is operated by the Smithsonian Astrophysical Observatory
for NASA under contract NAS8-03060.
M.\,B. acknowledges support from the International Fulbright Science and Technology Award.
The OVRO 40-m monitoring program is supported in part by NASA grants
NNX08AW31G and NNX11A043G, and NSF grants AST-0808050 and
AST-1109911. The National Radio Astronomy Observatory is a facility of
the National Science Foundation operated under cooperative agreement
by Associated Universities Inc. Support for CARMA construction was
derived from the states of California, Illinois, and Maryland, the
James S. McDonnell Foundation, the Gordon and Betty Moore Foundation,
the Kenneth T. and Eileen L. Norris Foundation, the University of
Chicago, the Associates of the California Institute of Technology, and
the National Science Foundation. Ongoing CARMA development and
operations are supported by the National Science Foundation under a
cooperative agreement, and by the CARMA partner universities. The \textit{Fermi} LAT Collaboration acknowledges generous ongoing support
from a number of agencies and institutes that have supported both the
development and the operation of the LAT as well as scientific data analysis.
These include the National Aeronautics and Space Administration and the
Department of Energy in the United States, the Commissariat \`a l'Energie Atomique
and the Centre National de la Recherche Scientifique / Institut National de Physique
Nucl\'eaire et de Physique des Particules in France, the Agenzia Spaziale Italiana
and the Istituto Nazionale di Fisica Nucleare in Italy, the Ministry of Education,
Culture, Sports, Science and Technology (MEXT), High Energy Accelerator Research
Organization (KEK) and Japan Aerospace Exploration Agency (JAXA) in Japan, and
the K.~A.~Wallenberg Foundation, the Swedish Research Council and the
Swedish National Space Board in Sweden.
Additional support for science analysis during the operations phase from the following agencies is also gratefully acknowledged: the Istituto Nazionale di Astrofisica in Italy and the Centre National d'\'Etudes Spatiales in France.

\footnotesize{
\bibliographystyle{mn2eb}
\bibliography{thbib}
}
\label{lastpage}

\end{document}